
\input harvmac

\Title{SPhT-91-189}{On Symmetries of Some Massless 2D Field Theories.}

\centerline{Denis BERNARD}
\centerline{Service de Physique Th\'eorique de Saclay
\foot{Laboratoire de la Direction des sciences de la mati\`ere
du Commissariat \`a l'\'energie atomique.}}
\centerline{F-91191, Gif-sur-Yvette, France.}
\bigskip \bigskip
\vskip 2.0 cm

We describe few aspects of the quantum symmetries of some massless
two-dimensional field theories. We discuss their relations
with recent proposals for the factorized scattering theories of the
massless $PCM_1$ and $O(3)_{\theta=\pi}$ sigma models. We use
these symmetries to propose massless factorized S-matrices for the
$su(2)$ sigma models with topological terms at any level, alias
the $PCM_k$ models, and for the $su(2)$-coset massless flows.


\def\bar{\overline}

\def\*{\star}
\def\({\left(}		
\def\){\right)}		
\def\[{\left[}		
\def\]{\right]}

%
%
\def\frac#1#2{{#1 \over #2}}		

\def\half{{1 \over 2}}
\def\d{\partial}

\def\2pi{\hbox{$2\pi i$}}

\def\dsl{\raise.15ex\hbox{/}\kern-.57em\partial}
\def\Dsl{\,\raise.15ex\hbox{/}\mkern-.13.5mu D}
%
%
\def\th{\theta}		
		\def\Ga{\Gamma}

\def\al{\alpha}
\def\ep{\epsilon}
\def\la{\lambda}	
\def\de{\delta}		\def\De{\Delta}
\def\om{\omega}

%
%
		\def\CC{{\cal C}}

\def\CJ{{\cal J}}		
\def\CM{{\cal M}}		
	\def\CQ{{\cal Q}}	
		\def\CU{{\cal U}}

%
%
\font\numbers=cmss12
\font\upright=cmu10 scaled\magstep1
\def\stroke{\vrule height8pt width0.4pt depth-0.1pt}
\def\topfleck{\vrule height8pt width0.5pt depth-5.9pt}
\def\botfleck{\vrule height2pt width0.5pt depth0.1pt}
\def\Zmath{\vcenter{\hbox{\numbers\rlap{\rlap{Z}\kern 0.8pt\topfleck}\kern
2.2pt
                   \rlap Z\kern 6pt\botfleck\kern 1pt}}}
\def\Qmath{\vcenter{\hbox{\upright\rlap{\rlap{Q}\kern
                   3.8pt\stroke}\phantom{Q}}}}
\def\Nmath{\vcenter{\hbox{\upright\rlap{I}\kern 1.7pt N}}}
\def\Cmath{\vcenter{\hbox{\upright\rlap{\rlap{C}\kern
                   3.8pt\stroke}\phantom{C}}}}
\def\Rmath{\vcenter{\hbox{\upright\rlap{I}\kern 1.7pt R}}}
\def\Z{\ifmmode\Zmath\else$\Zmath$\fi}
\def\Q{\ifmmode\Qmath\else$\Qmath$\fi}
\def\N{\ifmmode\Nmath\else$\Nmath$\fi}
\def\C{\ifmmode\Cmath\else$\Cmath$\fi}
\def\R{\ifmmode\Rmath\else$\Rmath$\fi}

\Date{12/91}

\def\gb{ {\bar g} }
\def\ysu{ {Y_{su(2)}} }
\def\bb{ $\bullet$ }

\lref\rZZ{A.B. and Al.B. Zamolodchikov, ``Massless factorized scattering
and sigma-models with topological terms", preprint ENS-LPS-335-1991.}
\lref\rWi{E. Witten, Commun. Math. Phys. 92 (1984) 455\semi
	A.M. Polyakov and P.B. Wiegmann, Phys. Lett. B41 (1984) 223.}
\lref\rAb{M.C.B. Abdalla, Phys. Lett. 152B (1985) 215}
\lref\rdeV{H. deVega, Phys. Lett. 87B (1979) 233.}
\lref\rLu{M. Luscher, Nucl. Phys. B135 (1978) 1.}
\lref\rDr{V.G. Drinfel'd, ``Quantum groups",
	Proc. of the ICM, Berkeley, (1986).}
\lref\rYang{D. Bernard, Commun. Math. Phys. 137 (1991) 191.}
\lref\rKZa{V.G. Knizhnik and A.B. Zamoldchikov, Nucl. Phys. B297 (1984) 83.}
\lref\rWa{M. Wakimoto, Commun. Math. Phys. 104 (1986) 605.}
\lref\rZam{Al.B. Zamolodchikov, Nucl. Phys. B358 (1991) 619.}
\lref\rZaZa{A.B. and Al.B. Zamolodchikov, Annals of Physics 120 (1979) 253.}
\lref\rKRS{P.P. Kulish, N.Yu. Reshestikhin and E.K. Sklyanin,
	Lett. Math. Phys. 5 (1981) 393. }
\lref\RSmi{F. Smirnov, ``Dynamical symmetries of massive integrable models",
preprint RIMS-772-838, 1991.}
\lref\rCoset{D. Kastor, E. Martinec and Z. Qiu,
Phys. Lett. 200B (1988) 434\semi
J. Bagger, D. Nemeshansky and S. Yankielowicz, Phys. Rev. Lett. 60 (1988) 389
\semi
F. Ravanini, Mod. Phys. Lett. A3 (1988) 397.}
\lref\rBL{D. Bernard and A. Leclair, Phys. Lett. 247B (1990) 309.}
\lref\rRSG{A. Leclair, Phys. lett. 320B (1989) 103\semi
	F. Smirnov, Int. J. Mod. Phys. A4 (1989) 4213\semi
	D. Bernard and A. Leclair, Nucl. Phys. B340 (1990) 721. }
\lref\rTBA{C.N. Yang, Phys. Rev. Lett. 19 (1967) 1312\semi
Al.B. Zamolodchikov, Nucl. Phys. B342 (1990) 695.}
\lref\rBaRe{V.V. Bazhanov and N.Yu, Reshetikhin, Prog. Theor. Phys.
Supplement No 102 (1990) 301\semi
N.Yu. Reshetikhin, Harvard preprint 1990. }
\lref\rSo{C. Crnkovic, G.M. Sotkov and M. Satnishkov, Phys. lett.
	226B (1989) 297.}
\lref\rNLC{D. Bernard and A. Leclair, ``Quantum group symmetries and
non-local currents in 2D QFT", preprint 10/90,
to appear in Commun. Math. Phys; and ``Non-local currents in 2D QFT: an
alternative to the QISM", Proc. of the ``Quantum Groups" conference,
Leningrad, Nov. 1990.}
\lref\rHaAf{F.D.M. Haldane, J. Appl. Phys. 57 (1985) 3359; I. Affleck,
Les Houches lecture 1988, Ed. E. Brezin and J. Zinn-Justin, North. Holl. 1990}
\lref\rGKO{P. Goddard, A. Kent and D. Olive, Phys. Lett. 152B (1985) 88.}
\lref\rCarg{D. Bernard, Cargese lectures 1991, preprint SPhT-91-124.}
\lref\rRS{N.Yu. Reshetikhibn and M. Semenov-Tian-Shansky, Lett. Math.
Phys. 19 (1990) 133.}

\noindent {\bf 0- Introduction.}

\noindent Factorized scattering theory \rZaZa~ usually applies
to two-dimensional massive integrable quantum field theory.
Recently, A.B. and Al.B. Zamolodchikov proposed scattering theories
for some massless models \rZZ. In the following, we present a re-reading
of their results based on the quantum symmetries of these models.
Even if they provide further arguments supporting their conjectures,
our comments do not form a proof
but they could be of some interests in relation with the problem of
solving the 2D integrable models from their quantum symmetries \rNLC \RSmi .
\bigskip

\noindent {\bf 1- The $PCM_k$ models.}

\noindent \bb {\it The classical action}. The principal chiral models
at level $k$, $PCM_k$, are sigmal models with topological terms.
The classical action is \rWi:
\eqn\eIi{ S_{PCM_{k}}\ =\ \frac{1}{4\la^2} \int d^2x\
tr\({\d_\mu g\, \d_\mu g^{-1}}\) + k\ \Ga(g) }
The topological term $\Ga(g)$ is defined by:
\eqn\eIii{ \Ga(g)\ =\ \frac{1}{24\pi} \int_B d^3y \ep^{ijk}
tr\({\gb^{-1}\d_i\gb\, \gb^{-1}\d_j\gb\, \gb^{-1}\d_k\gb }\) }
where $B$ is a three dimensional space whose boundary $\d B$ is
the two-dimensional space-time
and $\gb$ is an extension of $g$ on $B$. For the quantum theory
to be well-defined the parameter $k$ as to be an integer; it is called
the level. The classical equations of motion are:
\eqn\eIiii{\eqalign{
\d_\mu\ J^R_\mu\ &=\ 0 \qquad {\rm with}\qquad
J^R_\mu= g^{-1}\d_\mu g - \frac{k\la^2}{4\pi}\ep_{\mu\nu} g^{-1}\d_\nu g\cr
\d_\mu\ J^L_\mu\ &=\ 0 \qquad {\rm with}\qquad
J^L_\mu= \d_\mu g g^{-1} + \frac{k\la^2}{4\pi}\ep_{\mu\nu} \d_\nu g g^{-1}\cr}}
They are conservation laws for two left and right currents.

\noindent \bb {\it The classical integrability.} The $PCM_k$ are classically
integrable since they admit a zero curvature representation. Namely, the
classical equations of motion are equivalent to the vanishing of
the curvature, $[\d_\mu-A_\mu,\d_\nu-A_\nu]=0$, for the following
Lax connexion \rAb \rdeV:
\eqn\eIiv{
A_\mu =\ \({\frac{1+\frac{k \la^2}{4\pi} w}{1-w^2} }\)
\({ g^{-1}\d_\mu g + w \ep_{\mu\nu}g^{-1}\d_\nu g }\) }
where $w$ is a spectral parameter. Moreover, the
form of the Lax connexion immediately implies the existence of
an infinite number of non-local conserved currents. There exist
two infinite families of such currents: the first members of each of
these families are the left and right currents,  $J^L_\mu(x)$ and $J^R_\mu(x)$,
whereas the second ones, $\CJ^L_\mu(x)$ and $\CJ^R_\mu(x)$, are given by:
\eqn\eIv{\eqalign{
\CJ^L_\mu(x)\ &=\ \({1-(\frac{k\la^2}{4\pi})^2}\)\ep_{\mu\nu} \d_\nu gg^{-1}
+\half \[{ \int_{\CC_x} * J^L\ , J^L_\mu  }\] \cr
\CJ^R_\mu(x)\ &=\ \({1-(\frac{k\la^2}{4\pi})^2}\)\ep_{\mu\nu} g^{-1}\d_\nu g
+\half \[{ \int_{\CC_x} * J^R\ , J^R_\mu  }\] \cr}}
Here $\CC_x$ is a one-dimensional curve ending at the point $x$.
The conservation laws for these currents are simple consequences
of the classical equations of motion.

\noindent \bb {\it The quantum non-local currents.}
The currents \eIv~ can be quantized following the approach initialized
by L\"uscher \rLu ; they are
regularized using a point-splitting procedure. As it was argued
in ref. \rAb~, there are no anomalies and the regularized quantum currents
are still conserved; therefore the quantum $PCM_k$ models
are supposed to be integrable. These non-local
currents reflect Yangians symmetries \rDr . More precisely,
as in ref. \rYang ,
the global charges $Q^a$ and $\CQ^a$ associated to the currents
$J^{L,a}_\mu$ and $\CJ^{L,a}_\mu$ are identified as the generators of
an Yangian quantum algebra $Y(G)$. Similarly, the
global charges ${\bar Q}^a$ and ${\bar \CQ}^a$ corresponding to the currents
$J^{R,a}_\mu$ and $\CJ^{R,a}_\mu$ also generate an Yangian $Y(G)$.
Therefore, the $PCM_k$ models is invariant by $Y_L(G)\times Y_R(G)$.

\noindent \bb {\it The RG flows.} The one-loop beta function was computed
by Witten in ref. \rWi~ in the case where $G=su(N)$:
\eqn\eIvi{ \beta(\la,k)\ =\ -\frac{\la^2(N-2)}{4\pi}
\({1-\(\frac{k\la^2}{4\pi}\)^2}\) }
It indicates a RG fixed-point at $\la^2=\frac{4\pi}{k}$. This
point corresponds to the WZW models at level $k$ \rWi \rKZa .
The RG flows of the
$PCM_k$ models can be described as follows. The ultraviolet fixed
point is a free conformal field theory with central charge equal
to the dimension of the group, $c^{UV}=dim G$. The UV stress-tensor
can be written as :
\eqn\eIvii{
T^{UV}(z)\ =\ \half(i\d_z{\vec \phi})^2
- \sum_{\al>0} \om_{\al}^{\dag}\d_z\om_{\al}}
where $\vec \phi$ is a free boson field taking values in the Cartan subalgebra
of $G$ and the fields $\om_{\al}^{\dag}$, $\om_\al$, for any positive root
$\al$, form a system of spin 1-spin 0 conformal fields. For $k\not= 0$, the
infrared fixed point of the $PCM_k$ model is the $WZW_k$ conformal field
theory with central charge $c^{IR}=\frac{kdimG}{k+h^*}$, $h^*$ is the dual
Coxeter number \rKZa. Using a free field representation of the $WZW_k$ models
\rWa, the infrared stress-tensor can be written as:
\eqn\EIviii{
T^{IR}(z)\ =\ \half(i\d_z{\vec \phi})^2
- \frac{i}{\sqrt{k+h^*}}{\vec \rho}\cdot\d^2_z{\vec \phi}
- \sum_{\al>0} \om_{\al}^{\dag}\d_z\om_{\al} }
with ${\vec \rho}$ the Weyl vector of $G$.
The net effect of the RG flow on the stress-tensor is the production
of the background charge ${\vec \rho}/\sqrt{k+h^*}$ in
the IR regime and the
screening of degrees of freedom in accordance with the $c$-theorem
\foot{A WZW model can never be the ultraviolet fixed point of
a $G\times G$ invariant field theory.}.
We now restrict ourselves to the case $G=su(2)$.

\noindent {\bf 2- The Zamolodchikov's proposal for the S-matrix
of the $PCM_1$ models.}

\noindent Recently, A.B. and Al.B. Zamolodchikov proposed \rZZ~ factorized
scatterings for some integrable massless theories. Their strategy can be
summarized as follows. First they introduce a factorizable S-matrix for the
IR conformal field theory which acts separately on the left and right
sectors. Introducing the symbols $L_\al(\th)$ and $R_\al(\th)$ for the
chiral ``massless particles" with rapidities $\th$ the scattering theory
is then summarized by the following commutation relations \rZaZa :
\eqn\eIIi{\eqalign{
L_\al(\th_1)\ L_{\al'}(\th_2)\ &=\
S_{LL}(\th_1-\th_2)^{\beta\beta'}_{\al\al'}\
L_{\beta'}(\th_2)\ L_{\beta_1}(\th)\cr
R_\al(\th_1)\ R_{\al'}(\th_2)\ &=\
S_{RR}(\th_1-\th_2)^{\beta\beta'}_{\al\al'}\
R_{\beta'}(\th_2)\ R_{\beta}(\th_1)\cr }}
The analytical properties of the S-matrices $S_{LL}$ and $S_{RR}$ are
supposed to be the same as for massive scatterings. In particular,
besides satisfying the quantum Yang-Baxter equation, these S-matrices
are supposed to satisfy the unitary and crossing relations:
\eqn\eIIii{\eqalign{
&S_{12}(\th)\ S_{21}(-\th)\ =\ 1 \cr
&S_{12}(i\pi-\th)\ =\ S^{cross}_{12}(\th)
\equiv C_2 S_{12}^{t_2}(\th)C_2 \cr}}
Here $C$ is the charge conjugation matrix. We have used the standard
notation in which the lower indices refer to the spaces on which the
operators are acting.

The scattering theory for the massless (but not scale invariant)
theory is completed by specifying the scattering between the left and
the right particles:
\eqn\eIIiii{
L_\al(\th_1)\ R_{\al'}(\th_2)\ =\
S_{LR}(\th_1-\th_2)^{\beta\beta'}_{\al\al'}\
R_{\beta'}(\th_2)\ L_{\beta}(\th_1) }
The matrices $S_{LR}$ should satisfy mixed Yang-Baxter equations
for the consistency of the mixed scatterings $(LLR)$ or $(LRR)$.
As it was argued by Al.B. Zamolodchikov \rZam, the S-matrices $S_{LR}$
are not required to satisfy the unitary and crossing relations separately
but only a mixed unitary-crossing relation:
\eqn\eIIiv{
S_{12}(\th)\ S_{21}^{cross}(\th+i\pi)\ =\
S_{12}(\th)\ C_2 S_{21}^{t_2}(\th+i\pi)C_2\ =\ 1}

\noindent \bb {\it The S-matrix for the $WZW_1$ models.}
The S-matrix for the $su(2)$ WZW model at level one proposed in
ref. \rZZ~ can be described as follows: the chiral ``particles"
form $su(2)$ doublets, therefore we have the symbols $L_\al(\th)$
and $R_\al(\th)$ with $\al= +\ {\rm or}\ -$. The left-left and
the right-right scatterings was proposed to be:
\eqn\EIIv{ S_{LL}(\th)\ =\ S_{RR}(\th)\ =\ S_{\ysu}(\th) }
with \rZaZa~ :
\eqn\EIIvi{\eqalign{
S_{\ysu}(\th)^{\al'\beta'}_{\al\beta}\ &=\ u(\th)
\({\frac{ \th\de^{\al'}_{\al}\de^{\beta'}_{\beta} -
i\pi \de^{\beta'}_{\al}\de^{\al'}_{\beta} }{\th-i\pi} }\)\cr
u(\th)\ &=\ \prod_{n=1}^\infty \({
\frac{[i\pi(2n-2)+\th][i\pi(2n-1)-\th]}{[i\pi(2n-2)-\th][i\pi(2n-1)+\th]}
}\) \cr}}

\noindent \bb {\it The S-matrix for the $PCM_1$ models.}
Since the IR limit of the $PCM_1$ model is the $WZW_1$ model,
A.B. and Al.B. Zamolodchikov conjectured that the S-matrix of the
$PCM_1$ models is given by the $LL$ and $RR$ scattering defined in
eq. \EIIv~ plus a scattering between the left and right sectors.
By $su(2)$ invariance, the latter is purely diagonal and they find
\rZZ :
\eqn\EIIvii{ S_{LR}(\th)\ =\ T(\th)\
=\ \tanh\({\frac{\th}{2}-i\frac{\pi}{4} }\) }
The function $T(\th)$ is constrained by the unitary-crossing relation
\eIIiv : $T(\th)T(\th+i\pi)=1$. In ref. \rZZ , the authors choose the
``minimal" solution.

\noindent {\bf 3- Quantum symmetries of the $PCM_1$ models.}

\noindent \bb {\it Quantum symmetries of the $WZW_1$ models.}
First let us consider the $WZW_1$ models. It is well-known that
the S-matrix \EIIvi~ is $\ysu$ invariant;
on-shell the generators of $\ysu$ act on the particles
with rapidity $\th$ by $Q^a=t^a$ and $\CQ^a=i\frac{\th h^*}{2\pi}t^a$
where $t^a$ form the spin $\half$ representation of $su(2)$.
These $\ysu$ symmetries reflect the existence of the non-local currents
\eIv~ that we discussed in the previous section. Moreover, as we will show,
the $\ysu$ algebras can be recovered in the bootstrap approach,
or more precisely, they can be reconstructed
from the Zamolodchikov exchange algebra \eIIi
\foot{This construction arises from a joint work with A. Leclair on
properties of form factors in two dimensions. It will be further
developped elsewhere.}. The generators
of the quantum algebra appear in the fusion \rKRS~
of two spin $\half$ representations, since these representations are
self conjugate. Thus, following F. Smirnov \RSmi, we assume
that the OPE of two $L(\th)$ operators (or two $R(\th)$)
possesses a simple pole if the rapidities differ by $i\pi$, and
we introduce two generating matrices $T^R(\th)$ and $T^L(\th)$ by:
\eqn\EIIIi{\eqalign{
T^L_{\al\al'}(\th)\ &=\ {\rm Res}_{\th-\th'=i\pi}
\({ L_\al(\th)L_{\al'}(\th') }\) \cr
T^R_{\al\al'}(\th)\ &=\ {\rm Res}_{\th-\th'=i\pi}
\({ R_\al(\th)R_{\al'}(\th') }\) \cr}}
A simple computation using eq. \eIIi~ reveals that the matrix
$T^L(\th)$ satisfies quadratic commutation relations:
\eqn\EIIIii{\eqalign{
S_{21}(\th_2-\th_1)\ &T^L_1(\th_1)\
S_{21}(\th_2-\th_1-i\pi)\ T^L_2(\th_2)\cr
&=\ T^L_2(\th_2) S_{12}(\th_1-\th_2-i\pi)\
T^L_1(\th_1)\ S_{12}(\th_1-\th_2) \cr}}
Here the S-matrix is the $S_\ysu$ matrix, eq. \EIIv . A similar
equation holds for $T^R(\th)$.
The algebra \EIIIii~ is very similar but not identical to the
quantum affine algebra in the presentation formulated in ref.
\rRS. Furthermore, let us factorize $T^L(\th)$
as $T^L(\th)= \({t^-(\th)}\)^{-1} t^+(\th)$ with $t^+(\th)$
$\({t^-(\th)}\)$ regular at infinity (at the origin). A consistent
set of commutation relations for $t^+(\th)$ and $t^-(\th)$ are:
\eqn\EIIIiii{\eqalign{
S_{12}(\th_1-\th_2)\ t^\pm_1(\th_1) t^\pm_2(\th_2)\ &=\
t^\pm_2(\th_2) t^\pm_1(\th_1)\ S_{12}(\th_1-\th_2) \cr
S_{12}(\th_1-\th_2)\ t^+_1(\th_1) t^-_2(\th_2)\ &=\
t^-_2(\th_2) t^+_1(\th_1)\ S_{21}(\th_2-\th_1-i\pi) \cr}}
The first of these equations is one of the alternative presentation
of the Yangian $\ysu$, and their set defines a extension
of the Yangians. The same relation applies for the right sector.
In summary, applying this construction for both chiral sectors,
provides a way to reconstruct the $\ysu\times\ysu$ symmetry of
$WZW_1$ model in the bootstrap approach.

\noindent \bb {\it Symmetries of the $PCM_1$ models.}
As we argued in section 1,
the $PCM_1$ model is invariant under the algebra $\ysu\times\ysu$ in the
same way as the $WZW_1$ model possesses this invariance. Therefore, to
verify this assertion we have
to check that the scattering between the left and right sectors
does not spoil the independence between the left and right Yangians.
A simple computation indicates that the matrix $T^R(\th_1)$ and $T^L(\th_2)$
commute,
\eqn\EIIIiv{ T^L_1(\th_1) T^R_2(\th_2)\ =\ T^R_2(\th_2) T^L_1(\th_1) }
provided that the diagonal $LR$-scattering
matrix $T(\th)$ satisfies the unitary-crossing relation:
$T(\th)T(\th+i\pi)=1$. This shows that the $PCM_1$ are effectively
invariant under $\ysu\times\ysu$. It also gives another argument
supporting the Zamolochikov's S-matrix \EIIv -\EIIvii.

\noindent {\bf 4- Remark on the $O(3)_{\th=\pi}$ S-matrix.}

\noindent The $O(3)_\th$ models are 2D sigma models with target the
two-sphere and with a topological term whose coefficient is $\th$.
For $\th=0$ the theory is massive, while it has been argued that for
$\th=\pi$ it is massless with IR fixed point the $WZW_1$ model \rHaAf.
The RG flow arrives at the IR fixed point along the direction $J^a_L
{\bar J}^a_R$ where $J^a_L$ and ${\bar J}^a_R$ are the left and right
currents of the $WZW_1$ conformal field theory. Thus, the large
distance effective action is:
\eqn\eIVa{ S^{IR}_{ {O(3)_{\th=\pi}} }\ =\ S_{ {WZW_1} } +
\la' \int d^2x\ J^a_L{\bar J}^a_R }
The current-current perturbation \eIVa~ breaks the $\ysu\times\ysu$
symmetry of the $WZW_1$ model but it preserves its diagonal subalgebra.
More precisely, perturbatively the diagonal current $J^{diag}=J^L+J^R$
is conserved and curl-free. This implies a conservation law for a
non-local current $\CJ^{diag}$ whose definition is similar to those
introduced in eq. \eIv. However, due to the non-liniarity in \eIv,
the non-local current $\CJ^{diag}$ is not the sum of the $\ysu$ currents
$\CJ^L$ and $\CJ^R$ of the $WZW_1$ model but,
\eqn\eIVb{\eqalign{
J^{diag}_\mu\ &=\ J^L_\mu\ +\ J^R_\mu \cr
\CJ^{diag}_\mu\ &=\ \CJ^L_\mu +\CJ^R_\mu +
\half\[{\int_{{\CC_x}} *J^L, J^R_\mu}\] +
\half\[{\int_{{\CC_x}} *J^R, J^L_\mu}\] \cr}}
Therefore, for the global charges $Q^a_{diag}$ and $\CQ^a_{diag}$,
we have:
\eqn\eIVc{\eqalign{
Q^a_{diag}\ &=\ Q^a_L\ +\ Q^a_R \cr
\CQ^a_{diag}\ &=\ \CQ^a_L +\CQ^a_R
+ \half \ep^{abc} Q^b_L Q^c_R \cr}}
Eqs. \eIVc\ reflect the comultiplication in $\ysu$. In other words, the
diagonal subalgebra is defined through the comultiplication.

Assuming bindly that these perturbative arguments remain valid
non-perturbatively, we may conclude that the massless S-matrix of the
$O(3)_{\th=\pi}$ has to be $Y^{diag}_{su(2)}$ invariant. This constrains
the LR-scattering to be proportional to $S_\ysu$, and thus we get:
\eqn\eIVd{\eqalign{
S_{LL}(\th)\ =\ S_{RR}(\th)\ &=\ S_\ysu(\th) \cr
S_{LR}(\th)\ &=\ iS_\ysu(\th) \cr}}
This is exactly the Zamolodchikov's proposal \rZZ. (We choose the
proportionality coefficient in order to recover their result).

\bigskip
\noindent {\bf 5- A proposal for the $PCM_k$ S-matrices.}

\noindent \bb {\it Quantum symmetries of the $WZW_k$ models
and their S-matrices.}
As we already pointed out, the $WZW_k$ models are $\ysu\times\ysu$
invariant for any level $k$.  But for $k\geq 2$, the $WZW_k$ model
possesses an extra quantum symmetry which has been identified
as $\CU_q(su(2))\times\CU_q(su(2))$
with $q=-\exp\(\frac{-i\pi}{k+2}\)$ (more precisely it is a $RSOS$
version of it). The currents generating this symmetry are $\CJ_k(z)$
and ${\bar \CJ}_k({\bar z})$ \rCoset \rBL:
\eqn\EVi{ \CJ_k(z)\ =\ \frac{1}{k+4}\ \({J^a_{-1}\Psi^a}\)(z) }
where $J^a_{-1}$ is the $(-1)$-component of the WZW chiral current
and $\Psi^b(z)$ is the chiral primary field taking values in the
adjoint representation of $su(2)$. The conformal dimension of
this current is $\De_k=1+\frac{2}{k+2}$. The corresponding
charges $Q^{(k)}$ and ${\bar Q}^{(k)}$
have spin $\pm \frac{2}{k+2}$.  For $k=2$, $\CJ_{k=2}(z)$ is a
supersymmetric current and the quantum symmetry simply reduces
to a supersymmetry. By abuse of notation, we will refer to this
quantum symmetry as a $k^{th}$ fractional supersymmetry.

\def\sro#1{ S^{({#1})}_{rsos} }
In order to take into account of all symmetries, we are lead
to propose a S-matrix for the chiral $LL$ and $RR$ scattering
having a tensor product form. This proposal is also supported
by the knowledge of the S-matrix for the Thirring-like
massive perturbations of the WZW models \rBL. The chiral ``massless
particles" consist of doublets of kinks; i.e. we have the symbols
$L^\pm_{ab}(\th)$ and $R^\pm_{ab}(\th)$ with $a,b=0,\half,\cdots,
\frac{k}{2}$, $|a-b|=\half$ and where the indices $\pm$ refer to the
states in the spin $\half$ representation of $su(2)$.
These doublets of kinks are in one-to-one correspondence with the
conformal blocks of the chiral $WZW$ primary fields valued
in the spin $\half$ representation.
For their scattering we propose:
\eqn\EVii{ S_{LL}(\th)\ =\ S_{RR}(\th)\ =\
S_\ysu(\th)\otimes S^{(k)}_{rsos}(\th) }
The $S_\ysu$ matrix acts only the $\pm$ su(2) indices and its
expression is given in eq. \EIIvi. The $\sro{k}$ matrix
is the S-matrix of the $k^{th}$ restricted sine-Gordon models
\rRSG. It acts only on the kink indices $a,b ...$:
\eqn\EViii{
L^\al_{ad}(\th) L^{\al'}_{dc}(\th')\ =\ \sum_b\
\sro{k}(\th-\th')^{ab}_{dc}\
L^{\al'}_{ab}(\th') L^{\al}_{bc}(\th) }
with, $[a]=\sin\({(2a+1)\pi/(k+2)}\)$,
\eqn\EViv{\eqalign{
\sro{k}(\th)^{ab}_{dc}&=v(\th)\({\frac{[a][c]}{[d][b]}}\)^{\th/2i\pi}
\({ \sinh(\frac{i\pi-\th}{k+2})\de_{db} + \sinh(\frac{\th}{k+2})
\({\frac{[d][b]}{[a][c]}}\)^\half \de_{ac} }\) \cr
v(\th)\ &=\ \frac{1}{\sinh(\frac{i\pi-\th}{k+2})}\prod_{n=1}^\infty \({
\frac{\sinh(\frac{i\pi(2n-2)+\th}{k+2})
\sinh(\frac{i\pi(2n-1)-\th}{k+2})}{\sinh(\frac{i\pi(2n-2)-\th}{k+2})
\sinh(\frac{i\pi(2n-1)+\th}{k+2})}
}\) \cr}}
It satisfies the Yang-Baxter equation and the unitary and crossing relations,
\eqn\EVv{\eqalign{
\sum_n\ \sro{k}(\th)^{an}_{dc}\
\sro{k}(-\th)^{ab}_{nc}\ &= \de_{db} \cr
\sro{k}(i\pi-\th)^{ab}_{dc}&=
\sro{k}(\th)^{da}_{cb} \cr}}

By construction, the S-matrix \EVii~ is invariant under the
symmetry algebra $\({\ysu\times\CU_q(su(2))}\)_L\times
\({\ysu\times\CU_q(su(2))}\)_R$
\foot{ To reconstruct the $WZW$ models from their quantum symmetry
algebra was also proposed in ref. \rCarg, (section 4a, remark 3), but
obviously, the Zamolodchikov's approach are much more concrete.}.
The $\ysu$ invariance is explicit.
The action of the fractional supersymmetric charges $Q^{(k)}$
and ${\bar Q}^{(k)}$ on the kinks under which the S-matrix
$\sro{k}$ is invariant was described in ref. \rRSG.

Besides the arguments based on symmetries,
another check for the $WZW_k$ S-matrix \EVii~ is available via the
Thermodynamics Bethe Ansatz analysis \rTBA.  In ref. \rBaRe ,
among many other results, Bazahnov and Reshestikhin
have derived the TBA equations for the
massive models with S-matrix $S_\ysu\otimes\sro{k}$.
Therefore the UV limits of these TBA equations are those for
the massless scatterings \EVii. Hopefully, these limits were computed
in ref. \rBaRe: they find
$c=\frac{kdimG}{k+h^*}$ for the values of the central charges
of the UV conformal field theories, as it should be.

\noindent \bb {\it The S-matrices for the $PCM_k$ models.}
The infrared fixed point of the $PCM_k$ model is the $WZW_k$ theory.
The RG trajectory arrives at the infrared fixed point along the
direction of the field $\CJ_k.{\bar \CJ}_k$ \rKZa. The large
distance effective action of the $PCM_k$ model is therefore:
\eqn\EVvi{
S^{IR}_{{PCM_k}}\ =\ S_{{WZW_{k}}} + \la' \int d^2x\ \CJ_k.{\bar \CJ}_k }
Since the fractional supersymmetry generated by the currents
$\CJ_k$ and ${\bar \CJ}_k$ commute with the Yangians,
the perturbation \EVvi~ preserves the $\ysu\times\ysu$
symmetry of the $WZW_k$ model, as it should be.
However it breaks the $\CU_q(su(2))\times
\CU_q(su(2))$ symmetry. But, in the same way as for the $J^a{\bar J}^a$
perturbation discussed in section 4, because it is
a current-current perturbation for the
quantum algebra, it should preserve the diagonal $\CU_q(su(2))$
symmetry. Assuming that these perturbative arguments remain valid
non-perturbatively, we may conclude that the quantum symmetry of
the $PCM_k$ models is $\ysu\times\ysu\times\CU_q(su(2))_{diag}$.

Therefore, the simplest scattering matrix satisfying all these
symmetry properties is:
\eqn\EVvii{\eqalign{
S_{LL}(\th)\ =\ S_{RR}(\th)\ &=\ S_\ysu(\th)\otimes S^{(k)}_{rsos}(\th)\cr
S_{LR}(\th)\ &=\ T(\th)\otimes S^{(k)}_{rsos}(\th)\cr }}
$T(\th)$ is defined in eq. \EIIvii. It satisfies all the mixed
factorized equations for the $(LLR)$ or $(LRR)$ scattering since
$S_\ysu$ and $\sro{k}$ are solutions of the Yang-Baxter equation.
The unitary and crossing relations are also fulfilled; in particular,
the S-matrix $\sro{k}$ satisfies the RSOS version of the unitary-crossing
relation \eIIiv :
\eqn\EVviii{
\sum_n\ \sro{k}(\th)^{an}_{dc}\
\sro{k}(i\pi+\th)^{dc}_{an}\ = \de_{db} }
For $k=1$, the RSOS factors are absent and we recover the Zamolodchikov's
proposal. Notice also that, when reconstructring the quantum symmetry
generators as in section 3, the introduction a non-diagonal
RSOS scattering (for $k\geq2$) effectively breaks the
left$\times$right RSOS quantum symmetry, whereas the $\ysu\times\ysu$
symmetry is preserved since $T(\th)$ is scalar and satisfies the
unitary-crossing relation.

\noindent {\bf 6- A proposal for the $su(2)$-coset S-matrices.}

\noindent \bb {\it Symmetries of the $su(2)$-cosets and their S-matrices.}
The $su(2)$-coset models $\frac{su(2)_l\otimes su(2)_k}{su(2)_{l+k}}$ \rGKO,
which we denote by $\CM(l,k)$, $l>k$, are characterized \rCoset~ by
a non-local chiral symmetry algebra generated by two currents
$\Phi_l$ and $\Phi_k$:
\eqn\EVIi{
\Phi_l^{(l;k)}\ =\ \[{\frac{(l;adj)\otimes(k;\cdot)}{(l+k;\cdot)}}\]
\quad,\quad
\Phi_k^{(l;k)}\ =\ \[{\frac{(l;\cdot)\otimes(k;adj)}{(l+k;\cdot)}}\] }
Here the dot denotes the scalar representation. These fields exist
only if $k$ or $l \geq2$. Their conformal
dimensions are: $\De_l=1+\frac{2}{l+2}$ and $\De_k=1+\frac{2}{k+2}$.
The corresponding charges $Q^{(l)}$ and $Q^{(k)}$ have
spin $\frac{2}{l+2}$ and $\frac{2}{k+2}$.
They generate a fractional supersymmetry in the same way as the currents
\EVi\ did. In a Feigin-Fuchs representation of the $su(2)$ cosets \rCoset~,
the currents \EVIi~ differ from the currents \EVi~
only by the radius of compactification of the bosonic field,
up to a total derivative.

Once again, the simplest way to take care of both symmetries is
to have a S-matrix having a tensor product form.
We propose that the chiral ``massless particles" are kinks
carrying two couples of indices, one for each fractional supersymmetry:
$L_{a_l b_l;a_k b_k}(\th)$ and $R_{a_l b_l;a_k b_k}(\th)$,
with $a_l,b_l=0,\half,\cdots,l/2$, $|a_l-b_l|=\half$ and
similarly for $a_k,b_k$. And for their scattering:
\eqn\EVIii{
S_{LL}(\th)\ =\ S_{RR}(\th)\ =\
\sro{l}(\th)\otimes S^{(k)}_{rsos}(\th) }
The two factors act separately on the couples of indices as in eq. \EViii.
By construction it is invariant under two fractional supersymmetries
generated by charges with spins $\frac{2}{l+2}$ and $\frac{2}{k+2}$.

If $k=1$, the $\CM(l,k=1)$ models are the minimal conformal theories and
the corresponding factor is absent. In the limit $l\to\infty$ and $k$ fixed,
the $su(2)$-cosets models $\CM(l\to\infty,k)$ become the $WZW_k$ models;
in this limit, the S-matrix \EVIii~ flows to the $WZW_k$ S-matrix since
we have $\sro{l\to\infty}=S_\ysu$.

The work by Bazahnov and Reshestikhin \rBaRe~ provides a TBA check
for the S-matrix \EVIii. Indeed they also derived the TBA equations
for the massive theories with S-matrix $\sro{l}\otimes\sro{k}$.
The ultraviolet limits of these equations are those of the massless
models with scattering \EVIii. In these limits, they computed the
central charges of the UV conformal field theories and found
those of the $su(2)$ cosets.

\noindent \bb {\it The S-matrices for the $su(2)$-coset flows.}
The $su(2)$ coset flows are RG flows from the $\CM(l,k)$ models
to the $\CM(l-k,k)$ models \rSo. The UV fixed point $\CM(k,l)$
is perturbed by the relevent field
$\[{\frac{(l;\cdot)\otimes(k;\cdot)}{(l+k;adj)}}\]$, and the
IR limit approaches from the direction
$\[{\frac{(l-k;adj)\otimes(k;\cdot)}{(l;\cdot)}}\]$.
The small and large distance effective actions are:
\eqn\EVIv{\eqalign{
S^{UV}_{coset}\ &=\ S_{\CM(l;k)} + \la \int d^2x\
\[{\frac{(l;\cdot)\otimes(k;\cdot)}{(l+k;adj)} }\]\cr
S^{IR}_{coset}\ &=\ S_{\CM(l-k;k)} + \la' \int d^2x\
\[{\frac{(l-k;adj)\otimes(k;\cdot)}{(l;\cdot)} }\].\cr}}
Both the ultraviolet and the infrared perturbing fields are
scalar for the $su(2)$ level $k$. Neither of them
breaks the left and the right $k^{th}$ fractional supersymmetry.
Therefore, perturbatively the $su(2)$ coset models are
invariant under two independent left and right $k^{th}$ fractional
supersymmetry. On the other hand, the infrared perturbing
field is not scalar for the $su(2)$ at level $(l-k)$, but it
is a current-current perturbations, $\Phi^{(l-k;k)}_{l-k}
{\bar \Phi}^{(l-k;k)}_{l-k} $, for the $(l-k)^{th}$ infrared
fractional supersymmetry. Therefore it should preserve the
diagonal quantum symmetry. Taking this argument for granted
non-pertubatively leads to the following simple ansatz
for the S-matrix of the $su(2)$ cosets:
\eqn\EVIvi{\eqalign{
S_{LL}(\th)\ =\ S_{RR}(\th)\ &=\
\sro{l-k}(\th)\otimes S^{(k)}_{rsos}(\th)\cr
S_{LR}(\th)\ &=\
\sro{l-k}(\th)\otimes T(\th) \cr}}
It satisfies all the unitary, crossing and factorization relations.
In the limit $l\to\infty$, $(l-k)=n$ fixed, the $su(2)$  massless cosets
become the $PCM_n$ models: the S-matrices \EVIvi~ effectively
reduce to the scattering matrix \EVvii . Also in the case $l=2$,
$k=1$, which corresponds to the flow between the tricritical Ising
model to the Ising model, all the matrix element of \EVIvi~
becomes trivial except for the $LR$ scattering which turn out to
be scalar, $S_{LR}(\th)=T(\th)$, in agreeement with ref. \rZam.

\noindent {\bf 7- Conclusions.}

\noindent To support our proposals we only offered cross-arguments
and there is obviously a need for more checks. To derive the TBA
equations for the massless non-scale invariant scatterings \EVvii~
and \EVIvi~ could be one of the possible way to check them.
Another more challenging approach consists in reconstructing the
correlation functions for these massless models. In particular,
it could be interesting to apply this program to the conformal field
theories in order to test the idea that 2D integrable models can
be reconstructed from their quantum symmetries.

\bigskip \bigskip
\noindent {\bf Acknoledgements:} It is a pleasure to thank O. Babelon,
A. Leclair and F. Ravanini for discussions.

\listrefs

\end